\documentclass{article}

\usepackage[final,nonatbib]{neurips_2023}

\usepackage[utf8]{inputenc} 
\usepackage[T1]{fontenc}    
\usepackage{url}            
\usepackage{booktabs}       
\usepackage{amsfonts}       
\usepackage{nicefrac}       
\usepackage{microtype}      
\usepackage{xcolor}         
\usepackage{pdfpages}
\usepackage{hyperref}       
\usepackage{subcaption}
\usepackage[hyphenbreaks]{breakurl}

\title{Pre-training strategy using real particle collision data for event classification in collider physics}

\author{%
  Tomoe Kishimoto \\
  Computing Research Center,
  High Energy Accelerator Research Organization, KEK\\
  1-1 Oho, Tsukuba, Ibaraki, Japan \\
  \texttt{tomoe.kishimoto@kek.jp} \\
  \And
  Masahiro Morinaga \\
  International Center for Elementary Particle Physics, ICEPP, The University of Tokyo\\
  7-3-1 Hongo, Bunkyo-ku, Tokyo, Japan \\
  \texttt{morinaga@icepp.s.u-tokyo.ac.jp} \\
    \And
  Masahiko Saito \\
  International Center for Elementary Particle Physics, ICEPP, The University of Tokyo\\
  7-3-1 Hongo, Bunkyo-ku, Tokyo, Japan \\
  \texttt{saito@icepp.s.u-tokyo.ac.jp} \\
      \And
  Junichi Tanaka \\
  International Center for Elementary Particle Physics, ICEPP, The University of Tokyo\\
  7-3-1 Hongo, Bunkyo-ku, Tokyo, Japan \\
  \texttt{jtanaka@icepp.s.u-tokyo.ac.jp} \\
}

\begin{document}

\maketitle

\begin{abstract}
This study aims to improve the performance of event classification in collider physics by introducing a pre-training strategy. Event classification is a typical problem in collider physics, where the goal is to distinguish the signal events of interest from background events as much as possible to search for new phenomena in nature. A pre-training strategy with feasibility to efficiently train the target event classification using a small amount of training data has been proposed. Real particle collision data were used in the pre-training phase as a novelty, where a self-supervised learning technique to handle the unlabeled data was employed. The ability to use real data in the pre-training phase eliminates the need to generate a large amount of training data by simulation and mitigates bias in the choice of physics processes in the training data. Our experiments using CMS open data confirmed that high event classification performance can be achieved by introducing a pre-trained model. This pre-training strategy provides a potential approach to save computational resources for future collider experiments and introduces a foundation model for event classification.
\end{abstract}

\section{Introduction}
In collider physics, a significant number of events\footnote{The term ``event'' corresponds to ``image'' in image classification.} are produced from particle collisions using high-energy accelerators such as the Large Hadron Collider (LHC)~\cite{Evans_2008}. Event classification, which separates interesting signal events from background events, plays a crucial role in data analysis. Even though Deep Learning (DL) can provide significant discrimination power in this classification problem by exploiting its large parameter space, a large amount of data is necessary to maximize its performance. The training data are typically generated using Monte Carlo (MC) simulations based on signal and background process theories. Because there are many data analyses that target various signal events, such as Higgs boson measurements and new phenomena searches~\cite{atlas_summary}, preparing a large amount of training data using MC simulations for all analyses is computationally expensive. For example, next-generation LHC experiments require extensive computing resources for MC simulations~\cite{Collaboration:2802918}. Therefore, there is a need for a technique that maximizes DL performance even with a small amount of training data.

The transfer learning (TL) technique is a promising approach to reduce computational costs because it enables efficient training of target tasks even with a small amount of data~\cite{5288526}. The DL model comprises a stack of layers with nonlinear functions. The initial layers learn the local features of the data, and the subsequent layers learn the global features. This behavior indicates that the knowledge of the local features learned while solving a problem can be transferred and put to use to solve a different set of problems that involve the common local features. This is how the TL technique functions and helps reduce computational demand. In the present analysis workflow for collider physics, dedicated DL models for each data analysis are trained from scratch, indicating that a large amount of training data is required for each data analysis. Thus, the demand on computing resources for generating the training data can be saved when a pre-trained model that can be transferred to many data analyses could be built.

This study presents a strategy for building a pre-trained model for event classification and performance improvement. The CMS open data~\cite{refId0} from real particle collisions (hereafter called real data) collected by the CMS experiment~\cite{CMS:2008xjf} at the LHC were used in pre-training by employing a self-supervised learning technique. The remainder of this paper is organized as follows: Section~2 describes the related work, including the advantages of using real data. Section~3 summarizes the datasets used in the study. Section~4 provides details of the DL model and the proposed pre-training strategy. Section~5 presents the experimental results. Finally, Section 6 summarizes the findings of the study.

\section{Related work}
DL has been successfully adapted for event classification in collider physics. A previous study has reported that DL outperforms traditional machine learning methods, such as Boosted Decision Trees, by discovering powerful features and providing better discrimination power~\cite{Baldi:2014kfa}. Another study on transferability of DL models to different signal events has reported that DL provides discriminative power to other signals that vary kinematically~\cite{PhysRevD.101.035042}. Application of the TL technique for event classification using graph neural network architecture~\cite{https://doi.org/10.48550/arxiv.1806.01261} was investigated and found that it enabled us to to examine the transferability between event classifications with different numbers of reconstructed particles~\cite{Kishimoto:2022sS}. In contrast to the aforementioned works, this study introduces a novel approach that utilizes real data for pre-training. The advantages of using real data are as follows:

\begin{itemize}
 \setlength{\leftskip}{-15pt}
\item First, there is no need to generate a large amount of training data using MC simulations for pre-training, which saves computing resources.
\item Second, choice of the physics process for the MC simulation that is applied for pre-training can be arbitrary because it is assumed that the pre-trained model will be optimized for the chosen physics process. The bias of this choice is mitigated by using real data because many physics processes are included in the real data. This will ensure the transferability of pre-trained model for many data analyses.
\end{itemize} 
However, due to limited availability of true information in real data, in comparison to MC simulation data, an alternative training method based on a self-supervised technique was developed, as discussed in Section~\ref{sec:pretrain}. Building a pre-trained model using self-supervision has been actively discussed in other fields, such as natural language processing, as a foundation model~\cite{Bommasani2021FoundationModels}. This study introduces a similar idea for data analysis in collider physics.

\section{Datasets}
As mentioned in the previous sections, there are two phases of training in this study: pre-training and event classification. Details of the datasets for each phase are described below.

{\bf Pre-training dataset}:  The single electron and single muon datasets in the CMS open data~\cite{cms_electron, cms_muon}, which are collision events at a center of mass energy $\sqrt{s}$ = 13 TeV in 2015, were used in the pre-training.  The electron and muon are simply called lepton in this study. The following loose event selections were applied to the reconstructed particles (hereafter called objects\footnote{Collider physics terminologies are described in \cite{Collaboration_2008}}) in an event: 
\begin{itemize}
 \setlength{\leftskip}{-15pt}
\item There is at least one loose lepton with $p_{\rm T} >$ 10 GeV.
\item There are at least two $b$-jets with $p_{\rm T} >$ 20 GeV.
\item There are at least two light-jets with $p_{\rm T} >$ 20 GeV.
\end{itemize} 
These event selections were made with an aim to obtain events with a topology similar to that of the target event classification task. The selected events were split into approximately10$^{6}$, 10$^{5}$, and 10$^{5}$ events for the training, validation, and testing datasets, respectively.

{\bf Event classification dataset}: The training data were produced using simulations: the collision events were generated by MadGraph5\_aMC@NLO~\cite{Alwall:2014hca} at $\sqrt{s}$ = 13~TeV, showering and hadronization were performed by Pythia8~\cite{Sjostrand:2014zea}, and the detector response was simulated by Delphes~\cite{Selvaggi_2014}. Two-Higgs-doublet model~\cite{2HDM}, which introduces additional Higgs bosons, $H^{0}$, $A$ and $H^{\pm}$, was used as the signal event. The top-pair production of the Standard Model was used as the background event. The observed objects per event were one lepton, one missing transverse momentum (MET) due to undetected neutrino, two $b$-jets, and two light-jets. Total events generated independently for the training, validation, test phases for each signal and background process were approximately 5$\times$10$^{5}$, 5$\times$10$^{4}$, and 5$\times$10$^{4}$ events, respectively. 

Six objects in an event were used in this study: a lepton, a MET, two $b$-jets, and two light-jets. The four-momenta of each object ($p_{\rm T}$, $\eta$, $\phi$, mass) and object-type were used as input variables. The $\phi$ was converted to ($\sin\phi$, $\cos\phi$) to handle the periodicity correctly. The object-type was represented in a one-hot vector format: lepton, MET, $b$-jet, or light-jet. Log transformation was applied to $p_{\rm T}$ and mass to fit the values within a reasonable range. 
  
\section{DL model}
\begin{figure}[htbp]
 \includegraphics[width=1.0\textwidth]{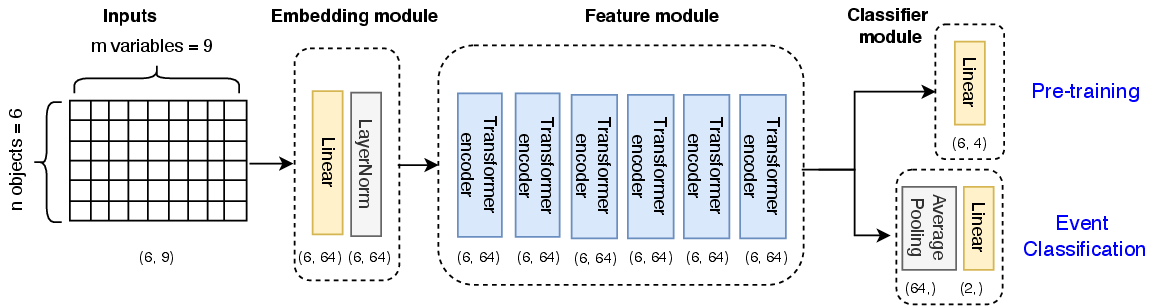}
 \caption{\label{fig:model} Overview of the proposed DL model. The numbers in the bracket indicate output shapes.}
\end{figure}

Figure~\ref{fig:model} shows an overview of the input data and the proposed DL model. The input data were prepared in a two-dimensional format of $n\times m$, where $n$ was the number of objects and $m$ was the number of feature variables. The model consisted of three modules: embedding, feature, and classifier modules. In the embedding module, the input feature variables for each object are embedded in a fully connected (linear) layer, and the outputs are then fed to the feature module. The transformer encoder layer~\cite{pytorch_trans} is employed in the feature module to exchange information between the objects, which is influenced by the natural language processing field. The feature module outputs are equivariant to permutations of objects. The classifier module consists of a linear layer and outputs predictions depending on the phase: pre-training or event classification. The total number of trainable parameters was approximately 1.7 M.

\subsection{Proposed pre-training strategy}
\label{sec:pretrain}
As mentioned earlier, a self-supervised learning technique was employed in the pre-training phase to handle unlabeled real data. In the pre-training phase, the object-type (lepton, MET, $b$-jets, or light-jets) in the feature variables was randomly masked by zeros when preparing a mini-batch: the probability of an object-type getting masked is 0.65. It was possible that all object-types in an event were masked or not masked at all. The DL model was trained to predict the masked object-types as a multi-label classification problem. All input feature variables, including the object-type, were used in the target event classification phase. The weight parameters of the embedding and feature modules, which were obtained by pre-training, were used as the initial parameters for event classification. These weight parameters were fine-tuned during the event classification phase. The classifier module was trained from scratch because it was assumed that the embedding and feature modules extracted common knowledge, and that the classifier module was highly dependent on the target task.

\section{Experiments}
The model proposed in this experiment was implemented using PyTorch~\cite{NEURIPS2019_9015} and is available in~\cite{hepfoundation}. The training settings were the same for the pre-training and event classification phases. The cross-entropy loss function was used as the loss function. Best epoch for the validation data is used as the final weight parameter after training for 100 epochs. The SGD algorithm~\cite{https://doi.org/10.48550/arxiv.1609.04747} was used as an optimizer, and the learning rate was decreased from 0.01 to 0.0001 by the cosine annealing algorithm~\cite{https://doi.org/10.48550/arxiv.1608.03983}. The batch size was set to 1,024. Other hyperparameters, such as the number of nodes and multi-heads in the transformer encoder layer, were optimized through a grid search using an event classification dataset without pre-training. All the executions used a local cluster of NVIDIA Tesla A100 cards, and training speed of approximately 90~batches/s was accomplished using an A100 card.

\begin{figure}[htbp]
  \begin{minipage}[b]{0.5\linewidth}
    \centering
    \includegraphics[clip, width=1\columnwidth]{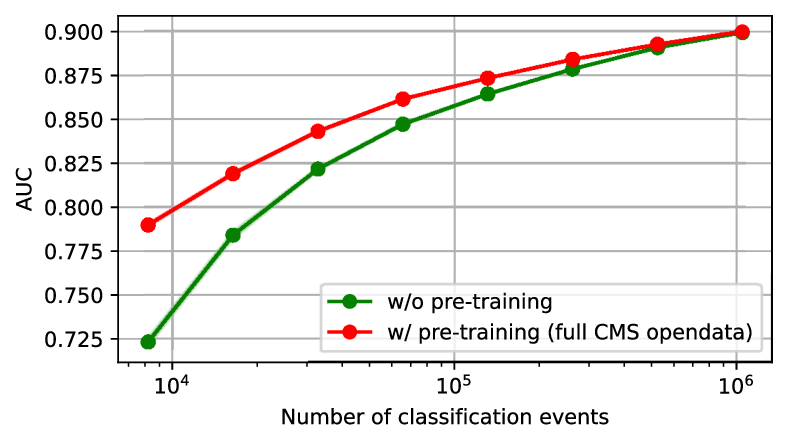}
        \subcaption{}
  \end{minipage}
  \begin{minipage}[b]{0.5\linewidth}
    \centering
    \includegraphics[clip, width=1\columnwidth]{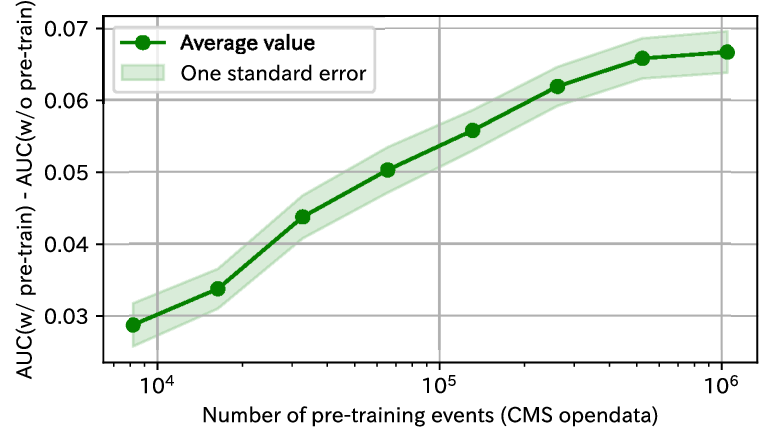}
            \subcaption{}
  \end{minipage}
    \caption{\label{fig:auc} (a) AUC values of the event classification task in terms of the number of classification events. The red and green points show the values with and without pre-training. (b) Differences of AUC values with and without pre-training in terms of the number of pre-training events. The points and error bands in both figures indicate the average value and one standard error of 50 runs. }
\end{figure}

Figure~\ref{fig:auc}~(a) shows the observed AUC values in the event classification task, with and without pre-training, that were obtained from the test dataset. The AUC values are shown in terms of the number of events used in the event classification phase, where all training events in the CMS open data were used for pre-training. 
Significant improvement in the performance could be evidenced by introducing pre-training when the number of events in the event classification was small (approximately 10$^{4}$ events). However, when the number of events increased to10$^{6}$ events and higher, it can be observed that the with and without pre-training performance curves converged. Furthermore, Figure~\ref{fig:auc}~(b) shows the AUC values with respect to the number of events used in the pre-training phase, where approximately 10$^{4}$ events were used in the event classification phase. It can be observed that the improvements achieved by introducing the pre-training phase increase as the number of events in the pre-training phase increases. 
 
Following have been identified as limitations related to this study. First, the scaling behavior shown in Figure~\ref{fig:auc}~(b) encourages a pre-training with a larger number of events; however, adding more events to the pre-training phase is difficult because the number of events available in the CMS open data itself is limited. Second, better optimization of the DL model and training strategy are required to achieve higher transfer learning efficiency. Furthermore, our interest lies in adapting the pre-trained model to different signal events to evaluate the generalization of the model. This will be the subject of future research.

\section{Conclusion}
In this study, a pre-training strategy that uses real data for event classification is discussed. The proposed model was successfully trained using real data by employing a self-supervised learning technique in which the object-types were masked for predictions. Our experiments confirmed that the AUC values in event classification could be improved by introducing a pre-trained model when the number of available events is small. Further, the experiments also indicated that improvements would be greater when more data were available in pre-training. In our experiment, event classification was performed using only one dataset. More physics processes must be investigated for event classification to improve the generalization of transferability, which is a subject for future research. 

This work can potentially contribute to reducing the demand for computing resources for future collider experiments because the need for generating a large amount of training data by simulation can be eliminated.
 This study aims to solve the problem of pure fundamental science, and we do not expect our study to result in a negative social impact.

\section*{Acknowledgment}
This work was supported by JSPS KAKENHI Grant Number JP22K14050.

\bibliographystyle{junsrt}
\bibliography{refs}

\appendix


\end{document}